\pdfoutput=1
\documentclass[12pt]{article}
\usepackage{algorithm}
\usepackage{algpseudocode} 
\usepackage{amsmath}
\usepackage{graphicx}
\usepackage{subcaption}
\usepackage{algpseudocode} 
\usepackage{float}
\usepackage{amssymb}
\usepackage{siunitx}
\PassOptionsToPackage{hyphens}{url}\usepackage{hyperref}
\usepackage{cleveref}
\usepackage[utf8]{inputenc}
\usepackage[right]{lineno}
\usepackage{csquotes}
\usepackage{booktabs}
\usepackage{longtable}
\usepackage{adjustbox}
\usepackage{array}
\usepackage{url}
\usepackage{titlesec}
\usepackage{authblk}
\usepackage{xcolor}
\usepackage{multirow}
\usepackage[numbers,sort&compress]{natbib}
\usepackage[margin=1in]{geometry}
\usepackage{setspace}
\title{Bayesian machine learning approach for recurrent events studies using Soft Bayesian Additive Regression Trees (SBART) }
\author{MengXing Chen$^{1,*}$ , Debajyoti Sinha$^{1}$, Antonio Linero$^{2}$ \\ $^{1}$Department of Statistics, Florida State University, Tallahassee, FL, USA. \\$^{2}$ Statistics and Data Sciences, The University of Texas at Austin, Austin, TX, USA. }

\begin{document}
\maketitle
\begin{abstract}
\begin{singlespace}{ 
Recurrent event data frequently arise in biomedical studies, where individuals may experience multiple recurrences of the same type of events, such as recurrent hospitalizations. This article introduces a nonparametric method for recurrent events under a Bayesian ensemble learning framework, called Soft Bayesian Additive Regression Trees (SBART) \citep{linero2018bayesian}, which combines multiple soft decision trees to achieve high predictive accuracy and a smooth estimator of the underlying intensity of the recurrent events. The proposed model represents the conditional intensity function of the non-homogeneous Poisson process as the product of a time-constant baseline, a subject-specific frailty random effect, and a nonparametric component capturing potentially nonlinear covariate effects and unknown interactions among covariates and time. A two-layer data augmentation scheme is employed to efficiently incorporate the SBART component within our computational algorithm. Simulation studies demonstrate that our method, called RecSBART in short, achieves superior accuracy in estimating cumulative intensity compared to existing approaches, even when our modeling assumptions are not true. With the Bayesian analysis of a study of recurrent hospitalizations of colorectal cancer patients, we further demonstrate our RecSBART method's ability to reveal and interpret the underlying complex relationships among covariates in a recurrent events study.}
\end{singlespace}
\end{abstract}
\newpage
\section{Introduction}
\label{s:intro}
Recurrent events arise when study subjects experience multiple recurrences of the same type of event over time, such as relapses or repeated hospitalizations. In many clinical studies, the usual aim is to assess the regression effects of multiple covariates whose interactions and nonlinear effects on the risk of recurrence may be unknown. For example, the rehospitalization study of \cite{gonzalez2005sex} includes 861 hospitalizations during January 1996 to June 2000 of 403 patients diagnosed with colorectal cancer. The covariates include gender, receipt of chemotherapy, tumor stage, and comorbidity. It is reasonable to consider that some covariates may exhibit temporal variation in their associations with the risk of recurrent hospitalizations. In addition, the presence of multiple covariates introduces multiple complexities, including unknown levels of interactions and non-linear covariate effects. These challenges may render any restrictive parametric model inadequate, necessitating the use of more flexible semiparametric approaches for recurrent events data.

 Widely used marginal approaches for analysis of recurrent events data (for example, \cite{andersen1982cox, lin2000semiparametric}) usually yield interpretable population-level estimates if the marginal regression structure is correctly specified. 
 However, their strict assumptions, such as a semiparametric regression function of the covariates, limit their ability to model the unknown complex relationships of the covariates with the recurrent event process. To avoid these assumptions for recurrent events data, \cite{murris2025random} presented a machine learning approach, RecForest, that applies the log-rank test to construct the tree structures that capture nonlinear and interaction effects of covariates on the marginal cumulative intensity function. Although stable and flexible, RecForest remains restricted to the estimation of the marginal risk and does not model and predict the subject-specific trajectories of events, which are often of interest in practice.

The shared frailty models directly address the limitations of marginal models by introducing a latent random effect, called a frailty, to capture the unobserved heterogeneity among subjects and the dependence within subjects  \citep{oakes1992frailty, sinha1993semiparametric,  cook2021independence}. Within this framework, the Bayesian implementation is appealing, particularly for methods relying on the proportional intensity models with frailty  \citep{sinha1993semiparametric, manda2005bayesian, pennell2006bayesian}. In these models, the recurrent events for each subject are usually represented as independent sample paths of a non-homogeneous Poisson process, with the intensity function exhibiting a proportional structure that incorporates a nonparametric baseline component, a log-linear covariate effect, and a multiplicative frailty. This Bayesian framework naturally supports probabilistic inference, uncertainty quantification, and straightforward estimation of the semiparametric intensity function and prediction of future recurrences. However, such models usually require covariate effects to remain constant over time, which may limit their applicability in settings where the relationships between covariates and time are more complex. 

For many statistical applications, various popular ensemble methods combine multiple weak learners to approximate complex regression functions, achieving improved performance and better generalization than individual base learners, and offering greater robustness overall \citep{dietterich2000ensemble, mienye2022survey}. Building on this idea, \cite{chipman2010bart} introduced Bayesian Additive Regression Trees (BART), a flexible and stable model for capturing complex functional relationships. While BART has shown strong performance in various domains, including causal inference \citep{hahn2020bayesian} and Bayesian time-varying regression coefficient models \citep{deshpande2026vcbart}, its use of binary decision trees can lead to non-smooth estimates. To overcome this, \cite{linero2018bayesian} proposed SBART, briefly described in Section \ref{s:re}, which replaces hard thresholding with probabilistic soft splits, allowing the regression model to adapt to unknown smoothness and reducing sensitivity to the curse of dimensionality. 
For recurrent events data, \cite{sparapani2021nonparametric} used BART to estimate the survival probability of the first event, without modeling the subject-specific intensity. \cite{basak2022semiparametric} extended SBART to clustered and interval-censored survival data. 
Further building on this, we develop an SBART-based method to estimate the intensity and related functions for the subject-specific recurrent event process, enabling the assessment of the risk of recurrence for each subject up to the terminal time.

In Section \ref{sec:main_section} of this paper, we propose a semiparametric method, RecSBART, where the recurrent events for each subject are assumed to follow a Nonhomogeneous Poisson process (NHPP), with the intensity function modeled as the product of a baseline term, a completely nonparametric function of time and covariates, and a subject-specific frailty to account for unobserved heterogeneity among subjects. 
To facilitate the associated Bayesian computation, we develop an efficient Markov chain Monte Carlo algorithm in Section \ref{sec:computation}  using a data augmentation step that leverages a thinned Poisson process \citep{adams2009tractable}.  In Section \ref{sec:modelling}, we present several representative simulation studies to demonstrate the advantages of our approach, yielding reasonable estimates of model quantities of interest even when the true data-generating mechanism is different from our model assumptions.   Section \ref{sec:application} applies our method to the Bayesian analysis of a study on recurrent rehospitalization of colorectal cancer patients. For our method with a nonparametric intensity function, we also present and demonstrate a new tool for effectively and practically assessing and summarizing the regression effects of a subset of covariates of interest and their interactions. Finally, Section \ref{sec:dis} summarizes the strengths and limitations of the proposed method and outlines directions for future work.

\section{Review of BART and SBART}
\label{s:re}
The BART framework introduced by \cite{chipman2010bart} is a flexible nonparametric tool that combines a large number of shallow decision trees (called weak learners) into an ensemble to achieve high predictive accuracy for 
$Y\sim N(f(\mathbf{x}),\sigma^2)$, where $\mathbf{x}$ is a $p$-dimensional covariate vector, and the unknown nonparametric regression function $f: \mathbb{R}^p \rightarrow \mathbb{R}$ is modeled as a sum $f(\mathbf{x}) = \sum\limits_{m=1}^{M}\, g(\mathbf{x}; \Psi_m,\varpi_m)$ of the outputs $g(\mathbf{x}; \Psi_m,\varpi_m)$ from $M$ regression trees $(\Psi_m,\varpi_m)$. Here $\Psi_m$ is the unknown decision tree structure and $\varpi_m =(\mu_{m1},\ldots,\mu_{s_m})$ is the set of unknown leaf node parameters of the decision tree $m$ with $s_m$ leaves. For identifiability, we adopt the default setting for $\sigma$ and set $\sigma^2 = 1$. The prior on $\Psi_m$ has two components that control tree's depth and structure. First, the probability of splitting further at each branch node decreases with depth $d$ as $\gamma(1+d)^{-\beta}$, where $d=0,1,\ldots$, $\gamma \in (0,1)$, and $\beta >0$. This discourages deep trees and reduces overfitting. Second, at a branch node $b$, the splitting rule takes the form $[x_l \leq C_b]$, where the variable $x_l$ is chosen uniformly from $\mathbf{x}=(x_1,\ldots,x_p)$, and the cut-point $C_b$ is sampled uniformly from the set of available splitting values. This means that $\mathbf{x}$ satisfying $x_l \leq C_b$ are assigned to the left child node, while the rest go to the right. The prior of $\varpi_m = (\mu_{m1},\ldots,\mu_{s_m})$ is $\mu_{ms} \overset{i.i.d.}{\sim} N(0,\sigma_\mu^2)$ for $s=1,\ldots,s_m$ and $m=1,\ldots,M$.

The output $g(\mathbf{x}; \Psi_m, \varpi_m)$ of the regression tree can also be written as the sum 
\begin{equation}\label{BART}
g(\mathbf{x}; \Psi_m, \varpi_m) = \sum\limits_{s=1}^{s_m}\, w_s(\mathbf{x}; \Psi_m) \mu_{ms}
\end{equation}
of ``weighted" values $\mu_{m1},\cdots,\mu_{ms_m}$, where $w_s(\mathbf{x}; \Psi_m)$ is an indicator function that assigns $\mathbf{x}$ to leaf node $s$, producing a piecewise-constant function with discontinuities at split boundaries. This limits BART because $g(\mathbf{x}; \Psi_m, \varpi_m)$ lacks smoothness, and reduces prediction and estimation performance when the unknown $f(\mathbf{x})$ is smooth. For  a continuous regression function $f(\mathbf{x})$, \cite{linero2018bayesian} introduced the SBART, which allows partial association among multiple leaf nodes via replacing the indicator function $w_s(\mathbf{x}; \Psi_m)$  in \eqref{BART} with a soft ``weight" function 
$\prod\limits_{b \in \mathcal{A}_m(s)}\, \psi(x_{l_b};C_b,\alpha_b)^{1-R_b} (1-\psi(x_{l_b}; C_b,\alpha_b))^{R_b}$,
where $\mathcal{A}_m(s)$ denotes the collection of ancestor nodes of leaf $s$, $R_b$ indicates whether the path to leaf $s$ goes right at node $b$, $\alpha_b$ is a bandwidth parameter controlling smoothness, and $\psi(x;c,\alpha)$ is the cumulative distribution function (CDF) of a location-scale family with location $c$ and scale $\alpha$. In this paper, $\psi(x; c,\alpha)$ is set to the inverse-logit function $\psi(x; c,\alpha) = [1+\exp(-(x-c)/ \alpha)]^{-1}$, with $\alpha$ controlling split softness. As $\alpha \rightarrow 0$, SBART reduces to the BART model. Additionally, \cite{linero2018bayesian} showed that SBART adapts automatically to unknown smoothness and sparsity levels. In the next section, we introduce our flexible semiparametric model for the intensity function of the recurrent events process,  as well as the associated Bayesian framework, RecSBART, with a two-layer data augmentation strategy to facilitate the posterior computation via an efficient MCMC algorithm.

\section{A Semiparametric Bayesian Model for Recurrent Events}
\label{sec:main_section}

Let $N_i(t)$ denote the number of recurrent events in the time interval $(0, t]$,  observed over $t \in (0,a_i]$, where $a_i$ is non-informative termination time for subject $i = 1, \ldots, n$ with fixed covariates $\mathbf{x}_i \in \mathbb{R}^p$. Given the  unobservable subject-specific frailty $W_i$ to account for the heterogeneity among subjects, we model $N_i(t)$ as a Non-Homogeneous Poisson Process (NHPP), expressed as $N_i(t) \sim \text{NHPP}(\lambda_i(t \mid W_i, \mathbf{x}_i))$,
with the conditional intensity function 
\begin{equation}\label{1}
    \lambda_i(t \mid W_i, \mathbf{x}_i) = \lambda_0 (t) W_i \Phi(b(t, \mathbf{x}_i))\ ,
\end{equation}
where $ \Phi(\cdot)$ is the cumulative distribution function (CDF) of the standard Normal distribution, $\lambda_0$ is the constant baseline intensity, and the nonparametric function  $b(t,\mathbf{x}): \mathbb{R}^{+} \times \mathbb{R}^p \rightarrow \mathbb{R}$ allows the time-varying, nonparametric covariate effects. For \eqref{1}, the baseline intensity function $\lambda_0(t)$ can be reduced to a constant $\lambda_0$ in the subsequent computation \citep{linero2022bayesian}. The unknown, possibly nonlinear effects of both $\mathbf{x}_i$ and $t$ are accommodated by the nonparametric function $b(t,\mathbf{x}_i)$.

Our model in \eqref{1} offers the following three major advantages. It avoids the assumptions of the semiparametric proportional intensity models \citep{sinha1993semiparametric, pennell2006bayesian} that have the restriction of $\frac{\lambda_i(t \mid W_i, \mathbf{x}_i)}{\lambda_k(t \mid W_k, \mathbf{x}_k)}$ being free of time $t$. The model in \eqref{1} reduces to a homogeneous Poisson process (HPP) when $b(t, \mathbf{x}_i)$ depends only on $\mathbf{x}_i$ and not on time $t$. As explained later in Section \ref{sec:computation}, the model in \eqref{2} simplifies computation under the thinned Poisson framework, since $\Phi(\cdot)\in(0,1)$ implies $\lambda_i(t\mid W_i,\mathbf{x}_i)\leq \lambda_0 W_i t$ for $t\in(0,a_i]$. 
Following \cite{hougaard1995frailty}, we assume the frailties $W_1,\cdots, W_n$ to have independent common Gamma density, $W_i \overset{i.i.d.}{\sim} \operatorname{Gam}(\eta,\eta)$, with  variance $\eta^{-1}$, $\mathbb{E}(W_i|\eta) =1$, and density $g(W_i \mid \eta) \propto W_i^{\eta-1}\exp{(-W_i\eta)}$. 
The restriction on the mean of $W_i$ is required to ensure model identifiability \citep{sinha1993semiparametric}.

We denote  the observed recurrent events data as \(\mathcal{D}=\{ 0<y_{i1}<\ldots<y_{in_i}< a_i;\mathbf{x}_i \in \mathbb{R}^p\}_{i=1}^n\), where $y_{i1}<\ldots < y_{in_i}$ are the ordered $n_i$ recurrence times for subject $i$ within the non-informative monitoring interval $(0,a_i)$.  The likelihood function $L(\mathcal{W},\Theta \mid \mathcal{D})$ of our model in \eqref{1}  is given by
\begin{equation}\label{2}
L(\mathcal{W},\Theta \mid \mathcal{D}) \propto
\prod_{i=1}^n
g(W_i \mid \eta)
\exp\bigr(
-\lambda_0 W_i \int_0^{a_i} \Phi\bigl(b(t,\mathbf{x}_i)\bigr)\,dt \bigr)
\left \{ \prod_{j=1}^{n_i}
\lambda_0 W_i \Phi\bigl(b(y_{ij},\mathbf{x}_i)\bigr)\right\},
\end{equation}
where $\mathcal{W}=(W_1,\ldots,W_n)$ has the joint density $\prod\limits_{i=1}^n\, g(W_i \mid \eta) $ based on the $\operatorname{Gam}(\eta,\eta)$ density $g(\cdot \mid \eta)$, and $\Theta=(\Psi,\varpi, \lambda_0,\eta)$. Here $(\Psi,\varpi)$ are the parameters associated with the unknown $b(t,\mathbf{x})$ which is modeled as the regression tree in \eqref{BART} with the ``soft" weight function $\omega_s(\mathbf{x},t;\Psi,\varpi)$.
To fully specify the model, we need the joint prior distribution $p(\Theta)$ of all the parameters of the RecSBART model in \eqref{1}. We assume that $p(\Theta) \propto p(\eta)\times p(\lambda_0)\times p(\varpi \mid \Psi)\times p(\Psi)$, where $p(\eta)$ is the prior density for the frailty precision $\eta\sim\operatorname{Gam}(a,b)$,
with fixed hyperparameters $a,b>0$ chosen to reflect the
subject-specific heterogeneity induced by $W_i$ in \eqref{1}, and $p(\lambda_0)$ is
the prior density for the baseline $\lambda_0\sim\operatorname{Gam}(\tilde a,\tilde b)$ with
known $\tilde a,\tilde b>0$.
The prior $p(\varpi\mid\Psi)$ is on the SBART leaf parameters
$\varpi=(\mu_{m1},\ldots,\mu_{ms_m})$ given the tree $\Psi$, and $p(\Psi)$ is the SBART tree
prior described in Section~\ref{s:re}.  
$p(\varpi \mid \Psi)$ is the prior for the leaf node values $\varpi=(\mu_{m1},\cdots,\mu_{ms_m})$, and $p(\Psi)$ represents the prior for the tree $\Psi$ of SBART as described in Section~\ref{s:re}. 
In this paper, we adopt the default regularized prior specification \citep{linero2018bayesian} for $\Psi=(\sigma_{\mu}, \gamma, \beta, r_{\alpha} )$ with $\sigma_{\mu}=\dfrac{3}{2 \sqrt{M}}$, $(\gamma, \beta)=(0.95,2)$ and $r_{\alpha} =10$. The bandwidth parameter $\alpha_b \sim \operatorname{Gam}(1,r_{\alpha})$ is shared across all branches within a single tree $\Psi$ in the SBART, allowing the model to flexibly adapt to varying levels of smoothness.

\section{Bayesian Computation via Data Augmentation}
\label{sec:computation}
 Obtaining posterior samples of $(\Theta)$ from the joint posterior $p(\Theta,\mathcal{W}\vert \mathcal{D})\propto L(\mathcal{W},\Theta\vert \mathcal{D})p(\Theta)$ with prior $p(\Theta)$ specified in last section is challenging because the evaluation of the likelihood contribution
of each subject $i$ in \eqref{2} involves the integral $\int_0^{a_i}\Phi(b(t,\mathbf{x})) dt$, causing an increase in the computational burden at each step of the MCMC algorithm. To overcome this challenge, we employ the data augmentation strategy of \cite{adams2009tractable}, which introduces a set of latent variables $\{G_{ik} \in (0,a_i],  k=1, \ldots, n_i^{*}\}$ sampled from an NHPP with the intensity $\lambda_0W_i(1-\Phi(b(t,\mathbf{x}_i))$. These $\{G_{ik};\,  k=1, \ldots, n_i^{*}\}$ are independent of the observed recurrence times $\{y_{ij}\}_{j=1}^{n_i}$, which are treated as event times from a thinned Poisson process. 
So, the complete data likelihood contribution given the augmented  latent $\{G_{ik}\}_{k=1}^{n_i^{*}}$ for subject $i$ is given by
\begin{equation}\label{4}
L^c(\mathcal{W},\Theta \mid \mathcal{D},\mathcal{G}) 
      \propto \prod_{i=1}^n g(W_i \mid \eta)\{\prod\limits_{j=1}^{n_i}\lambda_0W_i\Phi(b(y_{ij},\mathbf{x}_i))\} \{
    \prod\limits_{k=1}^{n^*_i}\lambda_0W_i(1-\Phi(b(G_{ik},\mathbf{x}_i)))\}e^{-\lambda_0W_ia_i}.
\end{equation}
Since the complete data likelihood in \eqref{4} is proportional to the likelihood 
\begin{equation}\label{BinaryL}
\prod_{i=1}^n \{\prod\limits_{j=1}^{n_i}\Phi(b(y_{ij},\mathbf{x}_i))\} 
\{\prod\limits_{k=1}^{n^*_i}(1-\Phi(b(G_{ik},\mathbf{x}_i)))\}  
\end{equation}
of binary response data as a function of $b$ , we can introduce another latent variable, $\mathcal{Z}=\{Z_{iq_i}, q_i=1,\ldots,n_i+n_i^*\}_{i=1}^n$ (following \citep{albert1993bayesian}) to obtain the augmented likelihood 
\begin{equation}\label{6}
\begin{aligned}
       \tilde{L} (\mathcal{W}, \Theta \mid \mathcal{D},\mathcal{G},\mathcal{Z}) & \propto \prod\limits_{i=1}^{n} g(W_i \mid \eta)  \{\prod\limits_{j=1}^{n_i}\lambda_0W_i\phi(z_{i(j+n^*_i)}|b(y_{ij},\mathbf{x}_i),1)  \times\\ &\prod\limits_{k=1}^{n^*_i}\lambda_0W_i\phi(z_{ik}|b(G_{ik},\mathbf{x}_i),1)\} \exp(-\lambda_0W_ia_i),
\end{aligned}
\end{equation}
where $\phi(x \mid \mu,\sigma^2)$ denotes the density of the Normal distribution with the mean $\mu$ and variance $\sigma^2$. For brevity, we will use the notation $p(\theta \mid \text{rest})$ to denote the conditional posterior of any parameters $\theta$ given the rest of the parameters and latent variables. For the frailties $ W_i $, the conditional posterior $p(W_i \mid \text{rest})$ is 
the Gamma density, $\operatorname{Gam}(\eta+n_i+n_i^*,\eta+\lambda_0a_i)$.
Similarly, the conditional posterior $p(\lambda_0 \mid \text{rest})$ of baseline $\lambda_0$ is $\operatorname{Gam}(\tilde{a}+\sum\limits_{i=1}^n(n_i+n_i^{*}),\tilde{b}+\sum\limits_{i=1}^n W_ia_i)$. The parameter $\eta$ with conditional posterior $p(\eta \mid \text{rest}) \propto  \frac{\eta^{n\eta+a-1}}{(\Gamma(\eta))^n}(\prod\limits_{i=1}^n W_i)^{\eta-1} \exp(-\eta(b+\sum\limits_{i=1}^nW_i))$ can be updated using slice sampling, which is well-suited for handling non-standard distributions and avoids the need to tune proposal distributions \citep{neal2003slice}. 
This approach generalizes the strategy previously developed for univariate survival models \cite{basak2022semiparametric, linero2022bayesian}. Algorithm \ref{alg:1} summarizes the steps of the data augmentation algorithm. We provide the full hierarchical model in the Web Appendix A.

\begin{algorithm}
\caption{Inference Algorithm}
\label{alg:1}
    \hspace*{\algorithmicindent} \textbf{Input:} Observed recurrent time points $y_{ij}$, initiated value of $\lambda_0$, $b(t,\mathbf{x})$, frailty effect $W_i$, shape parameter of the frailty density $\eta$ and SBART parameters, the observed data set $\mathcal{D}=\{ 0<y_{i1}<\ldots<y_{in_i}< a_i;\mathbf{x}_i \in \mathbb{R}^p\}_{i=1}^n$ \\
	\begin{algorithmic}[1]
\For{$\text{iter}=1,\ldots,K$}
    \For{$i=1,\ldots,n$}
        \State $p_i \sim \operatorname{Pois}(\lambda_0 W_i a_i)$
        \State $c_{im}\overset{i.i.d.}{\sim} U(0,a_i), \quad m=1,\ldots,p_i$ 

        \State $ u_{im}\overset{i.i.d.}{\sim} U(0,1), \quad m=1,\ldots,p_i$ 
        \State $\mathbf{G_i} \leftarrow 
        \{c_{im}:u_{im}\leq 1-\Phi(b(c_{im},\mathbf{x}_i)),\ m=1,\ldots,p_i\}
        =\{G_{i1},\ldots,G_{in_i^*}\}$ 
        \State $
        Z_{iq_i} \sim
        \begin{cases}
        N(b(G_{iq_i},\mathbf{x}_i),1)I(-\infty,0), 
        & q_i=1,\ldots,n_i^*,\\
        N(b(y_{i(q_i-n_i^*)},\mathbf{x}_i),1)I(0,\infty), 
        & q_i=n_i^*+1,\ldots,n_i+n_i^*.
        \end{cases}$ 
    \EndFor
    \State Update $\lambda_0$ from Equation \eqref{6}
    \State Update $W_i$, $i=1,\ldots,n$, from Equation \eqref{6}
    \State Update $b(\cdot)$ using SBART
    \State Update frailty parameter $\eta$ from Equation \eqref{6}
\EndFor
\end{algorithmic}
\end{algorithm} 

\section{Simulation Study}
\label{sec:modelling}
In this section, we conduct a simulation study to evaluate the performance of our proposed method, RecSBART, under various simulation models. Across these simulation models, we compare the performance of RecSBART with two existing approaches: (1) the semiparametric Bayesian method of \cite{sinha1993semiparametric} with the proportional intensity model, and (2) a frequentist tree-ensemble method, RecForest, proposed by \cite{murris2025random}, and implemented in the R package \texttt{recforest}.

We considered these simulation models (Simulation A-C), each using distinct intensity functions $\lambda_i(t,\mathbf{x}_i)$, as described below. Each simulated dataset has $n=200$ subjects, with $4$ covariates $\mathbf{x}_i=(x_{i1},x_{i2},x_{i3},x_{i4})$ where $x_{i1},x_{i2},x_{i3},x_{i4}\overset{i.i.d.}{\sim}U(0,1)$. Given the subject-specific intensity $\lambda_i(t;\mathbf{x}_i)$, recurrent event times were simulated from a Poisson process over the time window $(0,1]$. Each simulation model generates $K=20$ replicates of datasets. For the Bayesian methods used in the simulation study, we employ 2,500 burn-in iterations followed by 2,500 sampling iterations, and the tree-structured models are specified as an ensemble of 50 single decision trees.

\begin{itemize}

    \item \textbf{Simulation A} evaluates model stability under a Homogeneous Poisson Process (HPP), which violates the RecSBART model assumptions. The intensity function is defined as
    $$ 
    \lambda_i(t;\mathbf{x}_i) =2\exp(-\beta_i^{0.3})+W_i\quad
    \text{with} \quad W_i \sim U(0,1),
    $$
   where, in all three simulation settings, $ \beta_i = 0.25\sin(\pi x_{i1}x_{i2})+0.1(x_{i3}-0.5)^2+0.25x_{i4}.$
   
    \item \textbf{Simulation B} evaluates model stability under a Non-Homogeneous Poisson Process (NHPP), which violates the RecSBART model assumptions. The intensity function is defined as
    $$ 
    \lambda_i(t;\mathbf{x}_i) = 2\exp\left(-(\beta_i t)^{0.3}\right)+W_i 
    \quad \text{with} \quad W_i \sim U(0,1).
    $$ 
    
    \item \textbf{Simulation C} investigates estimation accuracy under a NHPP which correctly satisfies the RecSBART model assumptions. The intensity function is defined as
    $$ 
    \lambda_i(t;\mathbf{x}_i) = 2W_i \exp\left(-(\beta_i t)^{0.3}\right)\quad \text{with} \quad W_i \sim \operatorname{Gam}(20,20).$$
    
\end{itemize}

One approach to evaluate the goodness-of-fit for each of the models is to use the Martingale residuals,
\begin{equation}
\label{12}
\hat{M}_i(y_{ij}) = N_i(y_{ij} \mid \mathbf{x}_i) - \hat{\Lambda}_i(y_{ij}).
\end{equation}
For subject $i$, $N_i(y_{ij} \mid \mathbf{x})$ is the observed cumulative number of events up to $j$th recurrence time $y_{ij}$, $j=1,\ldots,n_i$, and $\hat{\Lambda}_i(y_{ij})=\int_{0}^{y_{ij}}\,\hat{\lambda}_i(t,\mathbf{x}_i)\,dt$ denotes the estimated cumulative intensity. For every simulated dataset, the Martingale residuals from the Bayesian method using the proportional intensity model have substantially higher variance compared to those from the RecSBART and RecForest methods (for one particular dataset under each simulation setting, we include plots of the empirical distributions of these $3$ residuals in the Web Appendix B). However, the patterns of the residuals obtained from RecForest are very similar to those from RecSBART, indicating comparable performance of both procedures in terms of goodness-of-fit.

To evaluate the precision and accuracy of model fitting, we use the Average Mean Squared Error, $\text{AMSE}(\hat{\Lambda}) = \mathbb{E}_{\mathbf{X}}[\int_0^1 ( \widehat{\Lambda}(t \mid \mathbf{X})-\Lambda(t \mid \mathbf{X}))^2\,dt]$ of the estimate $\widehat{\Lambda}(t \mid \mathbf{X})$ of the cumulative intensity $\Lambda(t \mid \mathbf{X})$, where $\mathbb{E}_{\mathbf{X}}$ is the expectation with respect to the simulation model. We approximate AMSE$(\widehat{\Lambda})$ as 
\begin{equation}\label{eq:MSE}
     \operatorname{AMSE}(\widehat{\Lambda}) \backsimeq \dfrac{1}{K} \sum\limits_{k=1}^{K}\,  MSE_k\ \text{with}\ \operatorname{MSE}_k=\dfrac{1}{n}\sum\limits_{i=1}^{n} \{\sum\limits_{g=1}^{20}(\widehat{\Lambda}_i(t_g\mid \mathbf{x}_i)-\Lambda_i(t_g \mid \mathbf{x}_i))^2 \Delta_g\}
\end{equation}
for each replicate $k=1,\ldots,K$ and $\Delta_g=(t_g-t_{g-1})$. We compute  
$\operatorname{MSE}_k$ using the equally spaced partition $0=t_0<t_1<\cdots <t_{20}=1$ of the time-interval $(0,1]$. In \eqref{eq:MSE}, $\widehat{\Lambda}_i(t_{g} \mid \mathbf{x}_i)$ denotes the estimate of the true $\Lambda_i(t_g \mid \mathbf{x}_i)$ of the observed subject $i$. 

The results of the $\operatorname{AMSE}(\hat{\Lambda})$ of all competing methods under three simulation models are summarized in Table \ref{table:12}, along with corresponding 95\%  Bootstrap Monte Carlo intervals. RecSBART achieved the smallest AMSE among the competing methods across all three simulation models, indicating high accuracy in estimating $\Lambda(t \mid \mathbf{x})$ even when the frailty distribution is misspecified. In particular, under Simulation B (NHPP, frailty misspecification), RecSBART attains an AMSE of 0.032 (95\% Monte Carlo interval : (0.031, 0.032)), compared to 0.036 (95\% Monte Carlo interval: (0.035, 0.036)) for RecForest. This difference is the largest among the three simulation settings, highlighting the robustness of RecSBART under frailty misspecification. We also found out that $\text{MSE}_k$ values from RecForest are more right-skewed compared to those from RecSBART.

\begin{table}[H]
\caption{The approximated expected (under the three simulation models) Average Mean Squared Error (AMSE) for RecSBART, RecForest, and the Bayesian method using the Proportional Intensity model is reported. Values in parentheses represent Bootstrap Monte Carlo intervals (2.5\%–97.5\%  quantiles). The methods with the lowest (and thus most desirable) values are shown in boldface.}
\begin{tabular}{lccc}
\hline
Methods                & \multicolumn{1}{l}{Simulation A}                                        & \multicolumn{1}{l}{Simulation B}                                        & \multicolumn{1}{l}{Simulation C}                                        \\ \hline
RecSBART               & \textbf{\begin{tabular}[c]{@{}c@{}}0.035\\ (0.034, 0.036)\end{tabular}} & \textbf{\begin{tabular}[c]{@{}c@{}}0.032\\ (0.031, 0.032)\end{tabular}} & \textbf{\begin{tabular}[c]{@{}c@{}}0.027\\ (0.026, 0.028)\end{tabular}} \\
RecForest              & \begin{tabular}[c]{@{}c@{}}0.036\\ (0.0357, 0.038)\end{tabular}         & \begin{tabular}[c]{@{}c@{}}0.036\\ (0.035, 0.036)\end{tabular}          & \begin{tabular}[c]{@{}c@{}}0.030\\ (0.029, 0.032)\end{tabular}          \\
Proportional Intensity & \begin{tabular}[c]{@{}c@{}}0.778\\ (0.740, 0.798)\end{tabular}          & \begin{tabular}[c]{@{}c@{}}1.035\\ (1.015, 1.041)\end{tabular}          & \begin{tabular}[c]{@{}c@{}}0.484\\ (0.419, 0.517)\end{tabular}          \\ \hline
\end{tabular}
\label{table:12}
\end{table}

To assess and compare the abilities of 3 competing methods to estimate the unobserved frailty $W_i$, we employ the $\operatorname{AMSE}(\widehat{W})$
approximated as 
$\operatorname{AMSE}(\widehat{W}) \backsimeq \dfrac{1}{K} \sum\limits_{k=1}^K \{\dfrac{1}{n}\sum\limits_{i=1}^{n}(\widehat{W}_i^{(k)}-W_i^{(k)})^2\}$, where $W_i^{(k)}$ is the true frailty of patient $i=1,\cdots,n$ of the simulated dataset $k=1,\cdots,K$. The value of the $\operatorname{AMSE}(\widehat{W})$ for RecSBART is 0.0447, compared to 0.0476 for the proportional intensity model, indicating that RecSBART method estimates the frailty $W_i$ better than the analysis under the proportional intensity model.


\section{Application: Recurrent Hospitalizations of Colorectal Cancer Patients}
\label{sec:application}
The study of recurrent hospitalizations since surgery of 403 colorectal cancer patients \citep{gonzalez2005sex} had a total of 458 recurrent hospitalizations. The dataset is available in the \texttt{frailtypack} package in R. This dataset has been widely analyzed in the literature \citep{gonzalez2005sex, rondeau2012frailtypack, khan2022accelerated}. However, these existing analyses do not explicitly account for potential interactions among covariates. In contrast, RecSBART can capture nonlinear covariate interactions and time-varying covariate effects. Later in this section, we use RecSBART to investigate evidence of complex interactions among the covariates in this dataset.

The main data analysis goal is to compare the risks of recurrence for patients with different values of $p=4$ potential explanatory variables, including gender (female or male), chemotherapy status (placebo or treatment), Dukes' stage (from 0 to 3) and Charlson's index (from 0 to 3). For convenience in the subsequent section, we denote gender as $X_1$ (categorical), chemotherapy status as $X_2$ (categorical), Dukes' stage as $X_3$ (continuous) and Charlson's index as $X_4$ (continuous).  

For RecSBART model, we specify a  $\operatorname{Gam}(40,0.5)$ prior for hyperparameter $\eta$. 
 Both RecSBART and RecForest are implemented using 50 trees. 
 We analyze the data using following three methods: (1) RecSBART, (2) RecForest \citep{murris2025random}, and (3) semiparametric Bayesian method under the proportional intensity model \citep{sinha1993semiparametric}. For the Bayesian methods, the MCMC tools use 2,500 burn-in iterations followed by 2,500 saved iterations. 

\begin{figure}[H]
    \centering
    \includegraphics[width=0.8\linewidth,height=0.55\linewidth]{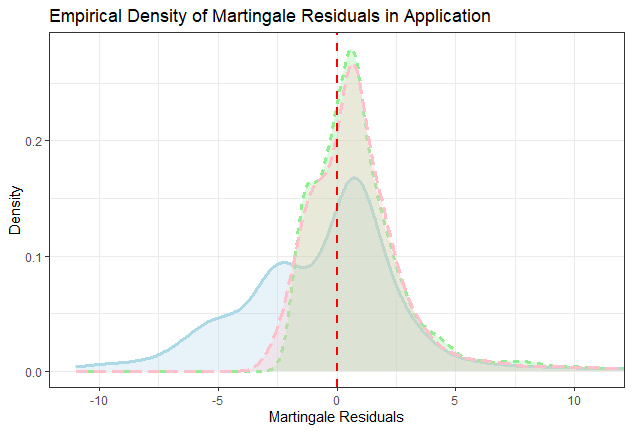}
    \caption{Empirical density of martingale residuals for RecSBART (with hyperparameter $\eta$ prior $G(40,0.5)$), RecForest, and the Bayesian method using the proportional intensity model in the rehospitalizations study. The light blue curve represents the Bayesian proportional intensity method, the light green curve represents the RecForest method, and the pink curve represents the RecSBART. This figure appears in color in the electronic version of this article, and any mention of color refers to that version.}
    \label{fig:three martingales in application}
\end{figure}

 
 We use the Martingale residuals defined in \eqref{12} to evaluate the goodness-of-fit of three competing methods. The empirical densities of the martingale residuals for all three methods are shown in Figure \ref{fig:three martingales in application}, where the RecSBART curve is more concentrated around zero and has lighter tails than those from the proportional intensity model and RecForest. Figure~\ref{fig:three martingales in application} clearly shows that the Bayesian method with proportional intensity has poorer goodness-of-fit compared to the others, but there is no clear winner between the remaining two methods. We compute the Mean Squared Martingale Residuals (MSMRs) on the full dataset to provide a quantitative evaluation for all three methods.
 \begin{equation}\label{msmr}
   \mathrm{MSMR}
=
\frac{1}{n}
\sum_{i=1}^{n}
\left\{
N_i(a_i) - \widehat{\Lambda}_{i}(a_i \mid \mathbf{x}_i)
\right\}^2,  
 \end{equation}
where $n$ is the total number of subjects, $N_i(a_i)$ denotes the observed cumulative number of recurrent events for subject $i$ up to terminal time $a_i$, and $\widehat{\Lambda}_{i}(a_i \mid \mathbf{x}_i)$ is the estimated cumulative intensity at $a_i$ given $\mathbf{x}_i$. The RecSBART attains the smallest value of 8.727, indicating superior overall fit, compared to 10.153 of RecForest and 14.514 of the Bayesian method with proportional intensity. To further assess generalization and potential overfitting, additional subject-level cross-validation results are provided in the Web Appendix C. The relative increase in MSMR from training to test is substantially larger for RecForest (28.4\%) than for RecSBART (8.7\%), indicating greater overfitting in RecForest.
 
 


\begin{figure}[H]
    \centering
    \includegraphics[width=0.75\linewidth,height=0.6\linewidth]{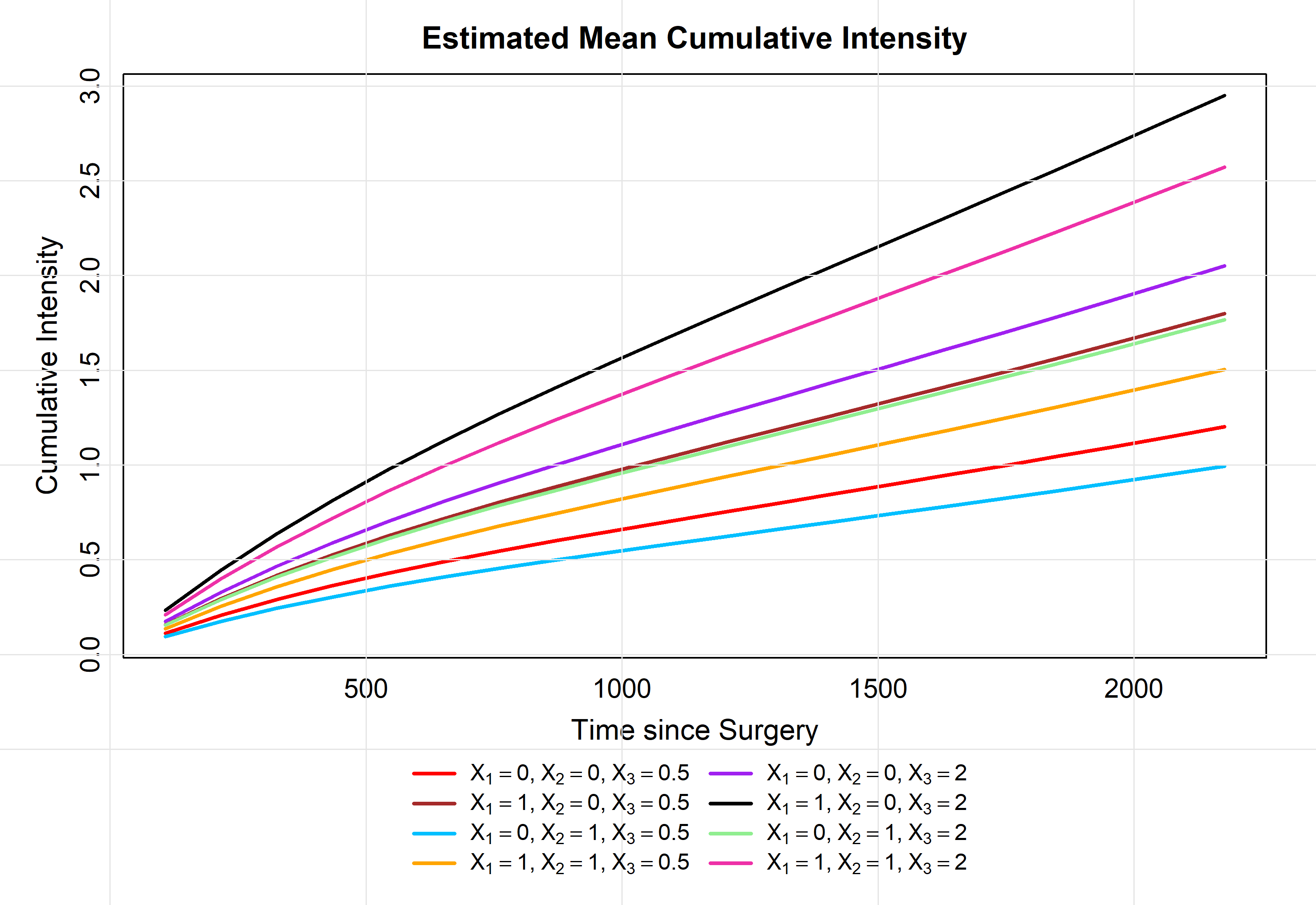}
    \caption{Estimated cumulative intensity $\hat{\Lambda}(t|\mathbf{x})$ of recurrent hospitalizations for patients with different combinations of gender ($x_1=0$ for female, $=1$ for male), chemotherapy status ($x_2=0$ for placebo, $=1$ for treatment), Dukes' stage ($x_3=0.5$ representing the 25$\%$ quantile , $=2$ representing the 75$\%$ quantile), and $x_4$ (Charlson's index) fixed at its observed median. This figure appears in color in the electronic version of this article, and any mention of color refers to that version.}
    \label{fig:estimate cumulative intensity in application}
\end{figure}

Insight into rehospitalization risk is obtained through the estimated cumulative intensity functions. $\hat\Lambda(t \mid \mathbf{x})$ of rehospitalization are estimated with combinations of gender ($x_1=0$ for female, $=1$ for male), chemotherapy status ($x_2=0$ for placebo, $=1$ for treatment), and Dukes' stage ($x_3=0.5$ representing the $25\%$ quantile, $=2$ representing the $75\%$ quantile), with $x_4$ (Charlson’s index) fixed at its median value. From Figure \ref{fig:estimate cumulative intensity in application}, males exhibit a higher cumulative intensity of recurrence (i.e., a higher risk of rehospitalization) than females at the same Dukes’ stage, regardless of chemotherapy treatment, which is consistent with the study of \cite{gonzalez2005sex}. Among patients receiving chemotherapy treatment, those with higher Dukes’ stage exhibit a higher risk of rehospitalizations, regardless of gender. Within the same gender, patients receiving placebo treatment only exhibit a higher risk of rehospitalization at the same Dukes' stage. This observation further motivates the study of interactions among covariates.

\begin{figure}[H]
    \centering

    \begin{subfigure}[t]{\textwidth}
        \centering
        \begin{subfigure}[t]{0.32\textwidth}
            \includegraphics[width=1.0\textwidth,height=1.1\textwidth]{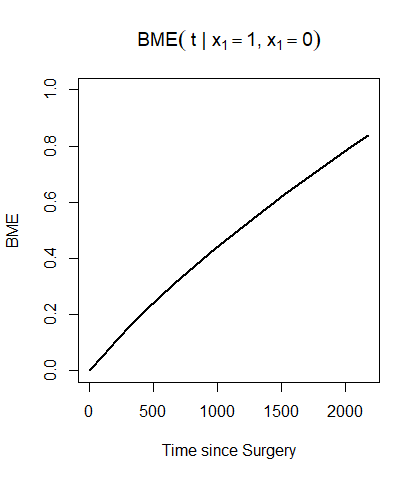}
        \end{subfigure}
        \hfill
        \begin{subfigure}[t]{0.32\textwidth}
            \includegraphics[width=1.0\textwidth,height=1.1\textwidth]{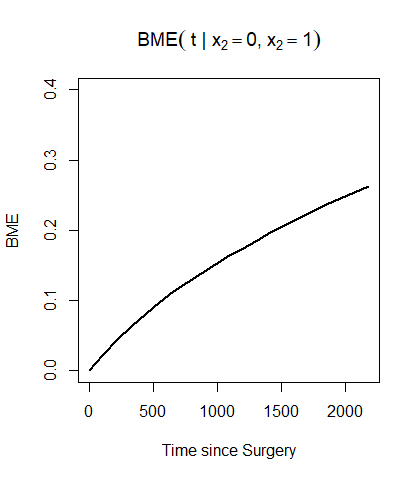}
        \end{subfigure}
        \hfill
        \begin{subfigure}[t]{0.32\textwidth}
            \includegraphics[width=1.0\textwidth,height=1.1\textwidth]{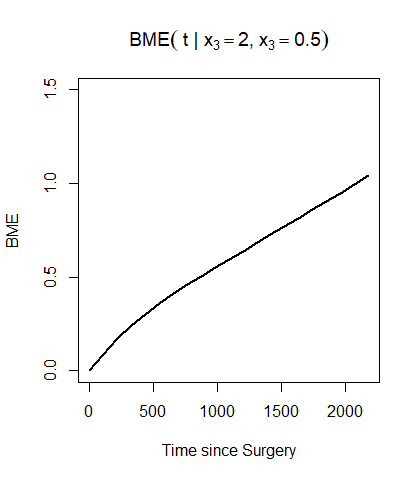}
        \end{subfigure}
        \caption{}
       \label{fig:single effect}
       
    \end{subfigure}
    
    \vspace{0.8cm}

    \begin{subfigure}[t]{\textwidth}
        \centering
        \begin{subfigure}[t]{0.32\textwidth}
            \includegraphics[width=1.1\textwidth,height=1.4\textwidth]{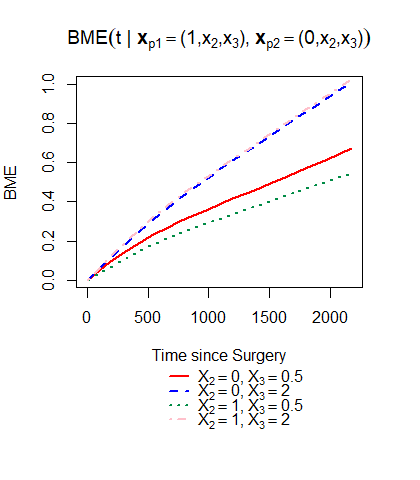}
        \end{subfigure}
        \hfill
        \begin{subfigure}[t]{0.32\textwidth}
            \includegraphics[width=1.1\textwidth,height=1.4\textwidth]{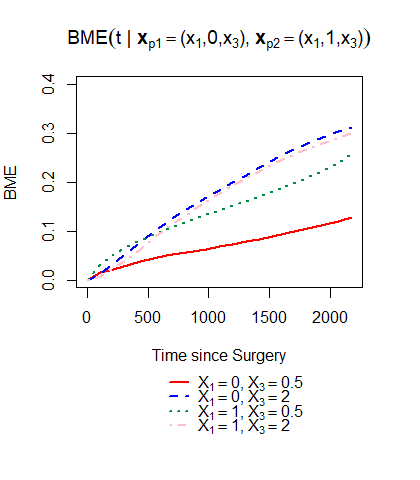}
        \end{subfigure}
        \hfill
        \begin{subfigure}[t]{0.32\textwidth}
            \includegraphics[width=1.1\textwidth,height=1.4\textwidth]{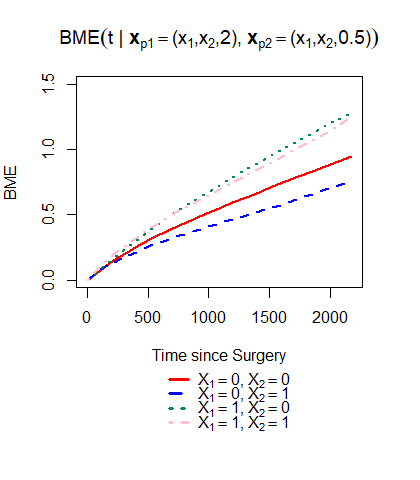}
        \end{subfigure}

        \caption{}
        \label{fig:interactions}
    \end{subfigure}

    \caption{Bayesian Marginal Effects (BMEs) in the rehospitalization study. The first row shows estimated marginal effects of the covariates $X_1$ (gender; $=0$ for female, $=1$ for male), $X_2$ (chemotherapy status; $=0$ for placebo, $=1$ for treatment), and $X_3$ (Dukes’ stage; $=0.5$ representing the 25$\%$ quantile , $=2$ representing the 75$\%$ quantile). The second row shows Bayesian estimates of the conditional effects of covariates $X_1$, $X_2$, $X_3$ given other two covariates. This figure appears in color in the electronic version of this article, and any mention of color refers to that version.}
    \label{fig:bme_combined}
\end{figure}

To quantify the magnitude of covariate effects, we introduce a Bayesian Marginal Effect (BME) that measures the effects of a subset of covariates, denoted by $\mathbf{X}_p$, inspired by \cite{dearmon2016local}. The BME is defined as
\begin{equation}\label{eq:BME}
    \operatorname{BME}(t\mid \mathbf{x}_{p1},\mathbf{x}_{p2})=\mathbb{E}_{\mathbf{X}_p^c}[\hat{\Lambda}(t \mid \mathbf{X}_p = \mathbf{x}_{p1},\mathbf{X}_p^c)-\hat{\Lambda}(t \mid \mathbf{X}_p = \mathbf{x}_{p2},\mathbf{X}_p^c)],
\end{equation}
as a measure of the average difference in prediction if all patients' $\mathbf{X}_p$ values  are changed from $\mathbf{X}_p=\mathbf{x}_{p1}$ to $\mathbf{X}_p=\mathbf{x}_{p2}$. By abuse of notation,  $\mathbf{X}_p^c$ denotes the remaining covariates,  
where $\mathbf{x}_{pi}^c$ for $i=1,\cdots, n$ are the observed values of $\mathbf{X}_p^c$ in the study. For illustrative purposes, we would like to use \eqref{eq:BME} to assess the regression effects and interactions of only $X_1$ (gender), $X_2$ (chemotherapy status), and $X_3$ (Dukes’ stage), and through only a few combinations of their values. In Figure~\ref{fig:single effect}, when the effects of the other two variables are ignored, these three BMEs increase over time, indicating that the risk of rehospitalization grows as time progresses. At the maximum observed time, the BME of $\mathbf{X}_p=X_1$ indicates that males experience 0.84 more rehospitalizations than females on average; the BME of $\mathbf{X}_p=X_2$ indicates that patients in placebo experience 0.26 more rehospitalizations than those receiving chemotherapy; and the BME of $\mathbf{X}_p=X_3$ indicates that patients with Dukes’ stage 2 experience 1.04 more rehospitalizations than those with lower stages. Subsequently, the BMEs are computed for different combinations of the covariate subset $\mathbf{X}_p = \{X_1, X_2, X_3\}$ to assess the interactions among these 3 covariates. As shown in Figure~\ref{fig:interactions}, when $X_3=2$ (the 75\% quantile of observed values), the BME values are similar across all combinations of gender ($X_1$) and chemotherapy status ($X_2$) indicating a weak interaction between gender and chemotherapy at the 75\% quantile (observed) of  $X_3$ (Dukes' stage). In contrast, when $X_3 = 0.5$ (the 25 \% quantile of observed values), the BME values are different across different combinations of gender and chemotherapy levels, suggesting a stronger interaction between these covariates at the 25\% quantile of Dukes' stage. When $X_1 = 1$ (male), the BME values remain almost unchanged across different combinations of Dukes' stage and chemotherapy level. However, when $X_1 = 0$ (female), the BME values vary across different combinations of Dukes' stage and  chemotherapy status, implying that the association between Dukes' stage and the outcome varies across chemotherapy status  among female patients. Overall, Figure~\ref{fig:bme_combined} suggests the presence of complex interactions among these $3$ variables.

To assess estimation uncertainty of the Bayesian marginal effects (BMEs), we employ a jackknife resampling procedure. Table~\ref{table:uncertainty} reports results for one representative covariate combination across three time points ($t=544, 1088, 1632$) corresponding to the first third, median, and last third of the observed time, where the corresponding 95\% Jackknife credible intervals are shown. For completeness, results under additional covariate combinations are provided in the Web Appendix D. The reported 95\% credible intervals are relatively narrow, indicating limited uncertainty in the BMEs estimates.
\begin{table}[H]
\centering
\caption{Bayesian marginal and conditional effects at selected time points for the colorectal cancer rehospitalization data. Effects are reported for $X_1$ (gender; 0 = female, 1 = male), $X_2$ (chemotherapy status; 0 = placebo, 1 = treatment), and $X_3$ (Dukes' stage; 0.5 = 25th percentile, 2 = 75th percentile). Values in parentheses are 95\% jackknife credible intervals based on the 2.5\% and 97.5\% quantiles.}
\label{table:uncertainty}
\begin{tabular}{llccc}
\toprule
Effect type & Covarite & $t=544$ & $t=1088$ & $t=1632$ \\
\midrule
Marginal
& $X_1$ & 0.2623 & 0.4742 & 0.6625 \\
&       & (0.2620, 0.2625) & (0.4737, 0.4746) & (0.6618, 0.6623) \\
& $X_2$ & 0.0971 & 0.1636 & 0.2170 \\
&       & (0.0970, 0.0972) & (0.1634, 0.1637) & (0.2167, 0.2172) \\
& $X_3$ & 0.3578 & 0.5959 & 0.8157 \\
&       & (0.3575, 0.3581) & (0.5952, 0.5964) & (0.8150, 0.8164) \\
\midrule
Conditional
& $X_1 \mid X_2=0, X_3=0.5$ & 0.2313 & 0.3869 & 0.5274 \\
&       & (0.2311, 0.2314) & (0.3866, 0.3871) & (0.5270, 0.5277) \\
& $X_2 \mid X_1=0, X_3=0.5$ & 0.0448 & 0.0684 & 0.0953 \\
&       & (0.0447, 0.0449) & (0.0682, 0.0685) & (0.0951, 0.0955) \\
& $X_3 \mid X_1=0, X_2=0$ & 0.3240 & 0.5521 & 0.7550 \\
&       & (0.3238, 0.3241) & (0.5517, 0.5523) & (0.7544, 0.7552) \\
\bottomrule
\end{tabular}
\end{table}

\begin{figure}[H]
    \centering

    \begin{subfigure}[t]{\textwidth}
        \centering
        \begin{subfigure}[t]{0.32\textwidth}
            \includegraphics[width=1.0\textwidth,height=1.1\textwidth]{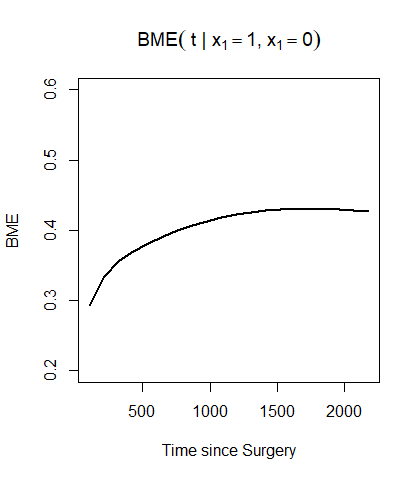}
        \end{subfigure}
        \hfill
        \begin{subfigure}[t]{0.32\textwidth}
            \includegraphics[width=1.0\textwidth,height=1.1\textwidth]{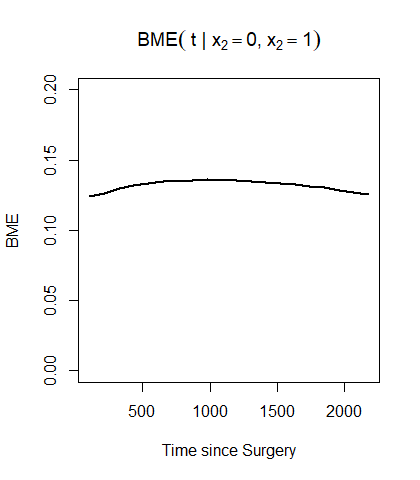}
        \end{subfigure}
        \hfill
        \begin{subfigure}[t]{0.32\textwidth}
            \includegraphics[width=1.0\textwidth,height=1.1\textwidth]{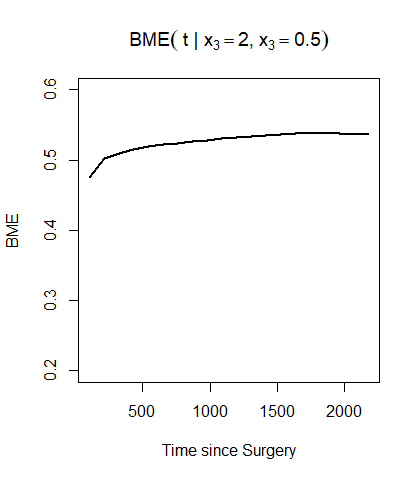}
        \end{subfigure}
        \caption{}
       \label{fig:log single effect}
       
    \end{subfigure}
    
    \vspace{0.8cm}

    \begin{subfigure}[t]{\textwidth}
        \centering
        \begin{subfigure}[t]{0.32\textwidth}
            \includegraphics[width=1.1\textwidth,height=1.4\textwidth]{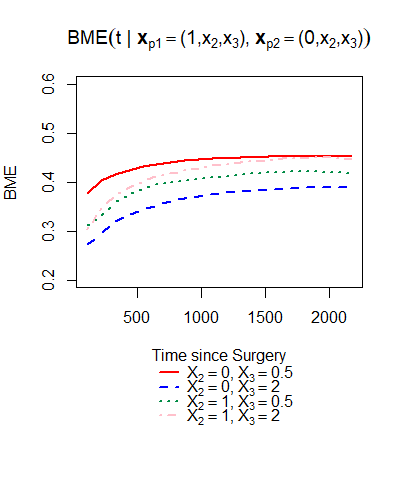}
        \end{subfigure}
        \hfill
        \begin{subfigure}[t]{0.32\textwidth}
            \includegraphics[width=1.1\textwidth,height=1.4\textwidth]{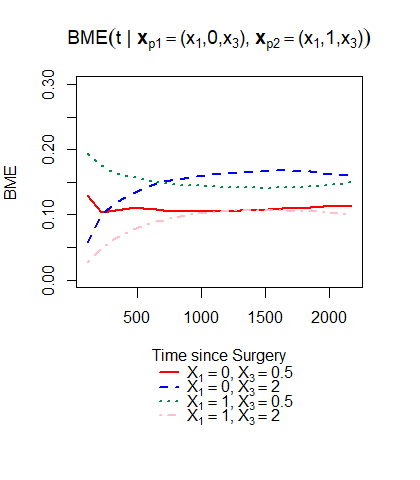}
        \end{subfigure}
        \hfill
        \begin{subfigure}[t]{0.32\textwidth}
            \includegraphics[width=1.1\textwidth,height=1.4\textwidth]{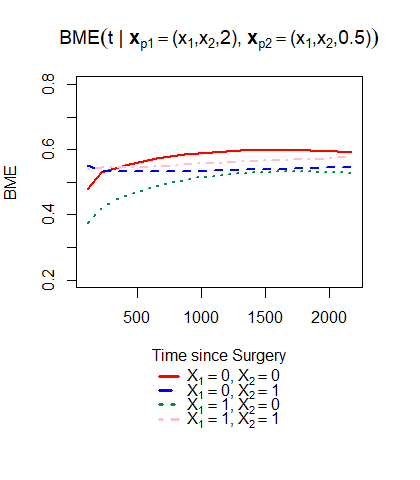}
        \end{subfigure}

        \caption{}
        \label{fig:log interactions}
    \end{subfigure}

    \caption{Bayesian Marginal Effects (BMEs) on the log cumulative intensity, $\log(\hat \Lambda(t \mid \cdot))$, in the rehospitalization study. The first row shows estimated marginal effects of the covariates $X_1$ (gender; $=0$ for female, $=1$ for male), $X_2$ (chemotherapy status; $=0$ for placebo, $=1$ for treatment), and $X_3$ (Dukes’ stage; $=0.5$ representing the 25$\%$ quantile , $=2$ representing the 75$\%$ quantile). The second row shows Bayesian estimates of the conditional effects of covariates $X_1$, $X_2$, $X_3$ given other two covariates. This figure appears in color in the electronic version of this article, and any mention of color refers to that version.}
    \label{fig:log-bme}
\end{figure}

We further employ BME on the log cumulative intensity function, $\log(\hat\Lambda(t\mid \mathbf{x}_{p1},\mathbf{x}_{p2}))$, which serves as a diagnostic for assessing the proportional intensity assumption because  the BME of $\log(\hat\Lambda(t\mid \mathbf{\cdot}))$ is expected to be constant over time $t$ if the proportional intensity assumption is true. 
We first computed BME values for the covariates $X_1$ (gender), $X_2$ (chemotherapy status), and $X_3$ (Dukes' stage) separately. As shown in Figure~\ref{fig:log single effect}, when the effects of other two covariates are ignored, $\text{BME}(t\mid x_1=1, x_1=0)$ exhibits a slight increase over time $t$, indicating a departure from the time-constant regression effect of $X_1$ on log-intensity. Figure~\ref{fig:log interactions} shows BME values  for different combinations of $X_p = \{X_1, X_2, X_3\}$. In particular, given $X_1 = 0$ (female) and $X_3 = 2$, the BME of $X_2$ increases gradually over time. These results suggest that a log-linear intensity model assumption for the covariate effects would have been unwise, because of the posterior evidence of the time-dependent covariate effects and possible interactions among covariates based on our analysis.

To assess the sensitivity of RecSBART to the choices of the different hyperparameters, we compare analysis results from using three prior distributions $\operatorname{Gam}(40,0.1)$, $\operatorname{Gam}(50,0.1)$, and $\operatorname{Gam}(70,0.5)$ for $\eta$.  The corresponding Mean Squared Martingale Residuals (MSMRs) defined in \ref{msmr} are 8.553, 9.120, and 8.802, respectively. The maximum relative difference among these metrics obtained using the four priors is 6.7$\%$, indicating that results from RecSBART are not sensitive to the prior choice for $\eta$.

\section{Discussion}
\label{sec:dis}
In this paper, we introduce RecSBART, a robust and flexible semiparametric method for recurrent events within the BART framework, which represents the underlying intensity of a non-homogeneous Poisson process as the product of a time-constant baseline,
a subject-specific frailty random effect, and a nonparametric component capturing potentially nonlinear covariate effects and unknown interactions among covariates and time. Using a two-layer data augmentation scheme, RecSBART enables efficient posterior computation. Through extensive simulation studies and an application to recurrent hospitalizations of colorectal cancer patients, we demonstrated that RecSBART provides accurate estimation of the cumulative intensity function and offers interpretable insights into underlying covariate–time relationships. Specifically, in Section \ref{sec:application}, the Bayesian Marginal Effects (BMEs) on the log cumulative intensity for covariates are not constant over time. This result indicated the presence of a complicated interaction between the covariates and time. Consequently, the proportional intensity assumption may be violated so that the proportional intensity model may be inappropriate for this study. 

We also report the computational cost of RecSBART under the current implementation. In Section \ref{sec:modelling}, Simulation B with 200 subjects and 348 recurrences required 0.12 s per iteration using unoptimized \texttt{R} code. In Section \ref{sec:application}, the algorithm required 0.21 s per iteration for 403 subjects with 458 recurrences in the rehospitalization study. All computations were performed on a desktop with a 13th Gen Intel (R) Core (TM) i5-13400F (2.50 GHz) processor and 16 GB RAM. The R code for RecSBART is available at: \url{https://github.com/mengxingchen1604/RecSBART}.  

One limitation of RecSBART is that it is currently restricted to vector-type covariates and responses, which limits its applicability to more complex data structures such as tensor  data. This restriction may prevent the model from fully capturing structured dependencies or correlations inherent in such data types. Future work may extend RecSBART to accommodate ultrahigh-dimensional covariates by incorporating sparsity-inducing priors, such as the Dirichlet prior proposed by \citet{linero2018bayesian}. Such extensions would be particularly useful when variable selection is a primary inferential goal.

\section*{Supplementary Material}
This supplementary material provides additional details on the hierarchical model specification, the goodness-of-fit metric, the overfitting assessment, and the uncertainty evaluation of the BMEs used in our study. It also presents additional simulation results that support the findings reported in the main text.

\subsection*{Web Appendix A}
In summary, our data augmentation scheme is based on the following hierarchical specification of the model we proposed. Each component of the specification including priors and likelihoods is modeled independently, reflecting modular assumptions about the underlying processes.
\begin{equation*}
    \begin{split}
        \text{prior:} \, & (\lambda_0,\eta,\Psi,\varpi) \sim p(\lambda_0)p(\eta)p(\varpi \mid \Psi)p(\Psi)\\
        & \lambda_0 \sim G(\tilde{a},\tilde{b})\\
        & \eta \sim G(a,b) \\
        \text{Model:} \,& W_i\mid \eta \overset{i.i.d.}{\sim} G(\eta,\eta), i=1,\ldots,n\\ \,& N_i(t) \mid W_i,\lambda_0,\Psi,\varpi \overset{i.i.d.}{\sim} \operatorname{NHPP}(\lambda_0W_i\Phi(b(t,\mathbf{x}_i))), i=1,\ldots,n \\
        & G_{ik} \mid W_i,\lambda_0,\Psi,\varpi \overset{i.i.d.}{\sim}\operatorname{NHPP}(\lambda_0W_i(1-\Phi(b(t,\mathbf{x}_i))), k=1,\ldots,n_i^*,i=1,\ldots,n\\
        & Z_{iq_i} \mid W_i,\lambda_0,\Psi,\varpi \overset{i.i.d.}{\sim}\begin{cases}
    N(b(G_{iq_i},\mathbf{x}_i),1)I(-\infty,0),& \text{if } q_i=1,...,n^*_i\\
   N(b(y_{i(q_i-n^*_i)},\mathbf{x}_i),1)I(0,\infty),& \text{if } q_i=n^*_i+1,...,n_i+n^*_i.
   \end{cases}
    \end{split}
\end{equation*}

\subsection*{Web Appendix B}
\begin{figure}[H]
    \centering
    \begin{subfigure}[t]{0.5\textwidth}
        \centering
        \includegraphics[width=1.2\textwidth, height=0.7\textwidth]{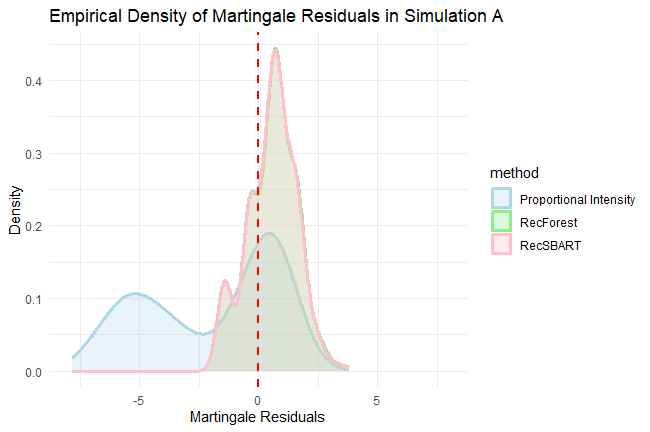}
        \caption{Simulation A}
        \label{fig:martA}
    \end{subfigure}
    \hfill
    \begin{subfigure}[t]{0.5\textwidth}
        \centering
        \includegraphics[width=1.2\textwidth,height=0.7\textwidth]{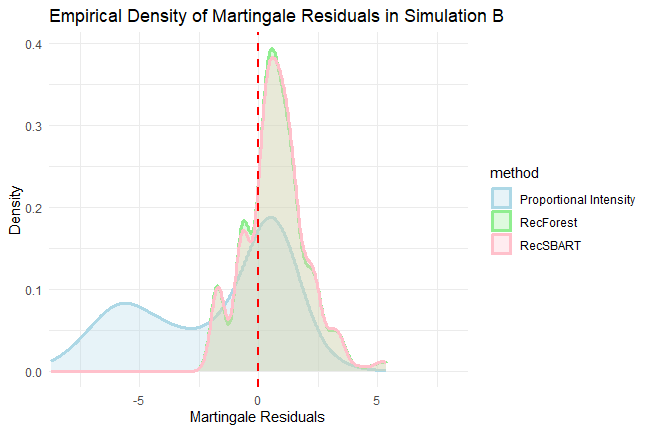}
        \caption{Simulation B}
        \label{fig:martB}
    \end{subfigure}
    \hfill
    \begin{subfigure}[t]{0.5\textwidth}
        \centering
       \includegraphics[width=1.2\textwidth,height=0.7\textwidth]{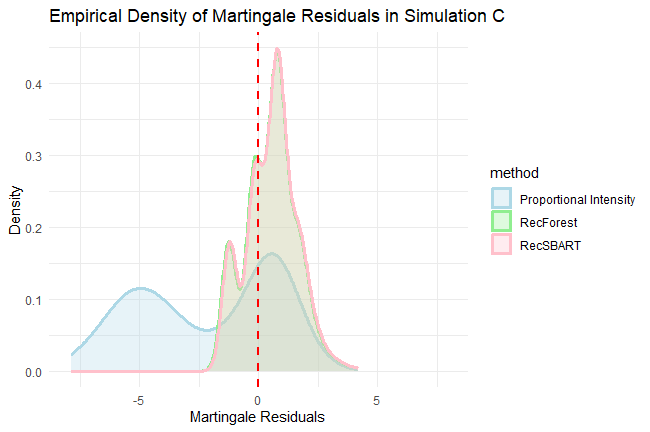}
        \caption{Simulation C}
        \label{fig:martC}
    \end{subfigure}

    \caption{Empirical densities of Martingale residuals under Simulation A–C for the three competing methods: RecSBART, RecForest, and the proportional intensity model. This figure appears in color in the electronic version of this article, and any mention of color refers to that version.}
    \label{fig:martingale simulation}
\end{figure}

To summarize the Martingale residuals as a single statistic, we compute the average of the sum of their absolute values, given by
\begin{equation*}
\label{13}
\frac{1}{K}\sum\limits_{k=1}^K(\sum_{i=1}^n\sum_{j=1}^{n_i}|\,\hat{M}^{(k)}_i(y_{ij})\,|),
\end{equation*}
where $\hat{M}^{(k)}_i(y_{ij})$ denotes the Martingale residual for subject $i$ at time $y_{ij}$ in the $k$th replicated dataset. The results are presented in Table \ref{11}. For Simulations A and C, the average sum of absolute Martingale residuals obtained from RecSBART is smaller than those from the two comparison methods, indicating that RecSBART achieves a better fit. In Simulation B, RecForest attains the lowest value, while RecSBART performs comparably, with only a modest increase in the residual sum. This behavior is expected because Simulation B satisfies the nonparametric assumptions underlying RecForest. As illustrated in Figure \ref{fig:martB}, the empirical Martingale residual distributions for both RecForest and RecSBART are more concentrated around zero and exhibit similar density shapes. In contrast, the residuals from the proportional intensity model display greater dispersion away from zero. Across all three simulation models, the Martingale residual analysis indicates that both RecSBART and RecForest provide superior model fit, whereas the proportional intensity model shows the poorest fit, reflecting its lack of robustness to assumption violations.

\begin{table}[H]

\caption{The approximate expected (under 3 simulation models) sums of absolute Martingale residuals for RecSBART, RecForest, and the Bayesian method using the proportional intensity model. Values in parentheses represent bootstrap percentile intervals (2.5th–97.5th percentiles). The methods with the lowest (most desirable) values are in boldface. }
\begin{tabular}{lccc}
\hline
Methods                & Simulation A                                                                  & Simulation B                                                                  & Simulation C                                                                  \\ \hline
RecSBART               & \textbf{\begin{tabular}[c]{@{}c@{}}468.995\\ (467.450, 470.416)\end{tabular}} & \begin{tabular}[c]{@{}c@{}}636.930\\ (0.031, 0.032)\end{tabular}              & \textbf{\begin{tabular}[c]{@{}c@{}}426.216\\ (425.406, 427.002)\end{tabular}} \\
RecForest              & \begin{tabular}[c]{@{}c@{}}474.563\\ (461.061, 487.991)\end{tabular}          & \textbf{\begin{tabular}[c]{@{}c@{}}630.326\\ (623.605, 636.794)\end{tabular}} & \begin{tabular}[c]{@{}c@{}}430.772\\ (424.531, 433.295)\end{tabular}          \\
Proportional Intensity & \begin{tabular}[c]{@{}c@{}}1233.017\\ (1184.017, 1281.695)\end{tabular}       & \begin{tabular}[c]{@{}c@{}}1390.472\\ (1359.027, 1423.205)\end{tabular}       & \begin{tabular}[c]{@{}c@{}}1239.354\\ (1239.791, 1246.615)\end{tabular}       \\ \hline
\end{tabular}
\label{11}
\end{table}

\subsection*{Web Appendix C} 
To compare potential overfitting in application between the two flexible methods, RecForest and RecSBART, we conducted 5-fold subject-level cross-validation. In each fold, one-fifth of the subjects were randomly assigned to the test set, and the remaining subjects were used for model training. The same folds were used for both methods to ensure a direct comparison. Predictive performance was evaluated using the Mean Squared Martingale Residual (MSMR). In fold $k$, this error was defined as
\[
\mathrm{MSMR}_{k}
=
\frac{1}{n_k}
\sum_{i \in \mathcal{T}_k}
\left\{
N_i(a_i) - \widehat{\Lambda}_{i}^{(-k)}(a_i \mid \textbf{x}_i)
\right\}^2,
\]
where $\mathcal{T}_k$ denotes the held-out subjects in fold $k$, $n_k = |\mathcal{T}_k|$, $N_i(a_i)$ is the observed cumulative number of recurrent events for subject $i$ up to terminal time $a_i$, and $\widehat{\Lambda}_{i}^{(-k)}(a_i \mid \textbf{x}_i)$ is the predicted cumulative intensity, trained without the subjects in fold $k$. We then summarized the cross-validated MSMRs across folds and compared them with the training MSMRs computed from the full dataset.

\begin{figure}[H]
    \centering
    \includegraphics[width=0.8\linewidth]{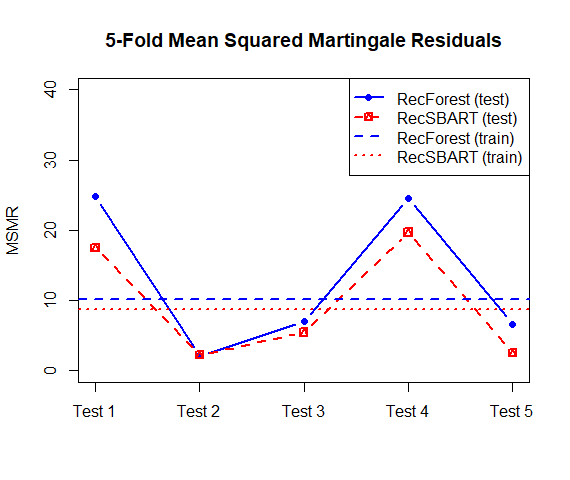}
    \caption{Five-fold subject-level cross-validated Mean Squared Martingale Residuals (MSMRs) for RecForest and RecSBART. Each point corresponds to the MSMR computed on a held-out test fold, while dashed horizontal lines indicate the corresponding training MSMRs obtained from the full dataset. This figure appears in color in the electronic version of this article, and any mention of color refers to that version.}
    \label{fig:5fold}
\end{figure}

As shown from the Figure~\ref{fig:5fold}, RecForest achieves a higher MSMR than RecSBART on every test set. This indicates that RecSBART consistently attains better out-of-sample predictive performance. Moreover, the variability of test MSMRs is substantially larger for RecForest, indicating greater sensitivity to the data split and a higher degree of overfitting. In contrast, RecSBART exhibits a smaller generalization gap and less variability across folds, indicating improved robustness in modeling the recurrent event process.

Table~\ref{table:overfitting}  summarizes the training and 5-fold cross-validated test MSMRs for RecForest and RecSBART. To quantify this difference, we report the relative gap, defined as
\[
\mathrm{Relative\ Gap}
=
\frac{\mathrm{MSMR}_{\mathrm{test}} - \mathrm{MSMR}_{\mathrm{train}}}
     {\mathrm{MSMR}_{\mathrm{train}}}.
\]
Under this measure, the test MSMR for RecForest exceeds its training counterpart by approximately 28.4\%, compared to only 8.7\% for RecSBART. This substantial difference indicates that RecForest  exhibits a higher overfitting.

\begin{table}[H]
\centering
\caption{Comparison of training and 5-fold cross-validated test performance for RecForest and RecSBART using the Mean Squared Martingale Residual (MSMR). Training MSMRs are computed on the full dataset, while test MSMRs are averaged across five subject-level folds.}
\begin{tabular}{llll}
\hline
Method    & MSMRs (train) & MSMRs (test) & Relative Gap \\ \hline
RecForest & 10.15       & 13.03      & 28.4\%        \\
RecSBART  & 8.73        & 9.49       & 8.7\%         \\ \hline
\end{tabular}
\label{table:overfitting}
\end{table}

\subsection*{Web Appendix D} 
For completeness, we report results under additional covariate combinations in this section. The reported 95\% credible intervals are relatively narrow, indicating limited uncertainty in the BME estimates.

\begin{table}[H]
\centering
\caption{Bayesian marginal effects under three representative covariate configurations. Effects are reported for $X_1$ (gender; 0 = female, 1 = male), $X_2$ (chemotherapy status; 0 = placebo, 1 = treatment), and $X_3$ (Dukes' stage). Values in parentheses are 95\% jackknife credible intervals based on the 2.5\% and 97.5\% quantiles.}
\label{table:combination_uncertainty}
\begin{tabular}{llccc}
\toprule
Combination & Covariate & $t=544$ & $t=1088$ & $t=1632$ \\
\midrule
Combination 1
& $X_1 \mid X_2=1, X_3=0.5$ & 0.1819 & 0.3120 & 0.4317 \\
&       & (0.1818, 0.1820) & (0.3118, 0.3121) & (0.4315, 0.4319) \\
& $X_2 \mid X_1=1, X_3=0.5$ & 0.0942 & 0.1433 & 0.1910 \\
&       & (0.0940, 0.0944) & (0.1430, 0.1436) & (0.1905, 0.1914) \\
& $X_3 \mid X_1=0, X_2=1$ & 0.2700 & 0.4360 & 0.5918 \\
&       & (0.2696, 0.2703) & (0.4354, 0.4364) & (0.5910, 0.5924) \\
\midrule
Combination 2
& $X_1 \mid X_2=1, X_3=3$ & 0.3232 & 0.5710 & 0.8003 \\
&       & (0.3231, 0.3234) & (0.5707, 0.5712) & (0.7998, 0.8005) \\
& $X_2 \mid X_1=1, X_3=3$ & 0.0850 & 0.1775 & 0.2485 \\
&       & (0.0848, 0.0852) & (0.1772, 0.1778) & (0.2481, 0.2489) \\
& $X_3 \mid X_1=1, X_2=1$ & 0.4113 & 0.6951 & 0.9603 \\
&       & (0.4109, 0.4117) & (0.6944, 0.6956) & (0.9600, 0.9610) \\
\midrule
Combination 3
& $X_1 \mid X_2=0, X_3=3$ & 0.3094 & 0.5640 & 0.7902 \\
&       & (0.3092, 0.3095) & (0.5637, 0.5642) & (0.7900, 0.7905) \\
& $X_2 \mid X_1=0, X_3=3$ & 0.0989 & 0.1845 & 0.2586 \\
&       & (0.0987, 0.0991) & (0.1842, 0.1848) & (0.2582, 0.2590) \\
& $X_3 \mid X_1=1, X_2=0$ & 0.4021 & 0.7292 & 1.0178 \\
&       & (0.4020, 0.4022) & (0.7288, 0.7294) & (1.0172, 1.0181) \\
\bottomrule
\end{tabular}
\end{table}
 \bibliography{reference}

\begin{thebibliography}{25}

\providecommand{\newblock}{}

\bibitem{linero2018bayesian}
Linero, Antonio R and Yang, Yun.
\newblock Bayesian regression tree ensembles that adapt to smoothness and sparsity.
\newblock \emph{Journal of the Royal Statistical Society Series B: Statistical Methodology}, 80(5):1087--1110, 2018. Oxford University Press.


\bibitem{gonzalez2005sex}
Gonz{\'a}lez, Juan Ramon and Fernandez, Esteve and Moreno, V{\'\i}ctor and Ribes, Josepa and Peris, Merc{\`e} and Navarro, Matilde and Cambray, Maria and Borr{\`a}s, Josep Maria.
\newblock Sex differences in hospital readmission among colorectal cancer patients.
\newblock \emph{Journal of Epidemiology \& Community Health}, 59(6):506--511, 2005. BMJ Publishing Group Ltd.


\bibitem{andersen1982cox}
Andersen, Per Kragh and Gill, Richard D.
\newblock Cox's regression model for counting processes: a large sample study.
\newblock \emph{The annals of statistics}, 1100--1120, 1982. JSTOR.


\bibitem{lin2000semiparametric}
Lin, Danyu Y and Wei, Lee-Jen and Yang, I and Ying, Zhiliang.
\newblock Semiparametric regression for the mean and rate functions of recurrent events.
\newblock \emph{Journal of the Royal Statistical Society: Series B (Statistical Methodology)}, 62(4):711--730, 2000. Wiley Online Library.


\bibitem{murris2025random}
Murris, Juliette and Bouaziz, Olivier and Jakubczak, Michal and Katsahian, Sandrine and Lavenu, Audrey.
\newblock Random survival forests for the analysis of recurrent events for right-censored data, with or without a terminal event.
\newblock 2025.


\bibitem{oakes1992frailty}
Oakes, David.
\newblock Frailty models for multiple event times.
\newblock \emph{Survival analysis: State of the art}, 371--379, 1992. Springer.


\bibitem{sinha1993semiparametric}
Sinha, Debajyoti.
\newblock Semiparametric Bayesian analysis of multiple event time data.
\newblock \emph{Journal of the American Statistical Association}, 88(423):979--983, 1993. Taylor \& Francis.


\bibitem{cook2021independence}
Cook, Richard J and Lawless, Jerald F.
\newblock Independence conditions and the analysis of life history studies with intermittent observation.
\newblock \emph{Biostatistics}, 22(3):455--481, 2021. Oxford University Press.


\bibitem{manda2005bayesian}
Manda, Samuel OM and Meyer, Renate.
\newblock Bayesian inference for recurrent events data using time-dependent frailty.
\newblock \emph{Statistics in medicine}, 24(8):1263--1274, 2005. Wiley Online Library.


\bibitem{pennell2006bayesian}
Pennell, Michael L and Dunson, David B.
\newblock Bayesian semiparametric dynamic frailty models for multiple event time data.
\newblock \emph{Biometrics}, 62(4):1044--1052, 2006. Oxford University Press.


\bibitem{dietterich2000ensemble}
Dietterich, Thomas G.
\newblock Ensemble methods in machine learning.
\newblock \emph{International workshop on multiple classifier systems}, 1--15, 2000.


\bibitem{mienye2022survey}
Mienye, Ibomoiye Domor and Sun, Yanxia.
\newblock A survey of ensemble learning: Concepts, algorithms, applications, and prospects.
\newblock \emph{Ieee Access}, 10:99129--99149, 2022. IEEE.


\bibitem{chipman2010bart}
Chipman, Hugh A and George, Edward I and McCulloch, Robert E.
\newblock BART: Bayesian additive regression trees.
\newblock 2010.


\bibitem{hahn2020bayesian}
Hahn, P Richard and Murray, Jared S and Carvalho, Carlos M.
\newblock Bayesian regression tree models for causal inference: Regularization, confounding, and heterogeneous effects (with discussion).
\newblock \emph{Bayesian Analysis}, 15(3):965--1056, 2020. International Society for Bayesian Analysis.


\bibitem{deshpande2026vcbart}
Deshpande, Sameer K and Bai, Ray and Balocchi, Cecilia and Starling, Jennifer E and Weiss, Jordan.
\newblock VCBART: Bayesian trees for varying coefficients.
\newblock \emph{Bayesian Analysis}, 21(1):281--308, 2026. International Society for Bayesian Analysis.


\bibitem{sparapani2021nonparametric}
Sparapani, Rodney and Spanbauer, Charles and McCulloch, Robert.
\newblock Nonparametric machine learning and efficient computation with Bayesian additive regression trees: The BART R package.
\newblock \emph{Journal of Statistical Software}, 97:1--66, 2021.


\bibitem{basak2022semiparametric}
Basak, Piyali and Linero, Antonio and Sinha, Debajyoti and Lipsitz, Stuart.
\newblock Semiparametric analysis of clustered interval-censored survival data using soft Bayesian additive regression trees (SBART).
\newblock \emph{Biometrics}, 78(3):880--893, 2022. Wiley Online Library.


\bibitem{adams2009tractable}
Adams, Ryan Prescott and Murray, Iain and MacKay, David JC.
\newblock Tractable nonparametric Bayesian inference in Poisson processes with Gaussian process intensities.
\newblock \emph{Proceedings of the 26th annual international conference on machine learning}, 9--16, 2009.


\bibitem{linero2022bayesian}
Linero, Antonio R and Basak, Piyali and Li, Yinpu and Sinha, Debajyoti.
\newblock Bayesian survival tree ensembles with submodel shrinkage.
\newblock \emph{Bayesian Analysis}, 17(3):997--1020, 2022. International Society for Bayesian Analysis.


\bibitem{hougaard1995frailty}
Hougaard, Philip.
\newblock Frailty models for survival data.
\newblock \emph{Lifetime data analysis}, 1(3):255--273, 1995. Springer.


\bibitem{albert1993bayesian}
Albert, James H and Chib, Siddhartha.
\newblock Bayesian analysis of binary and polychotomous response data.
\newblock \emph{Journal of the American statistical Association}, 88(422):669--679, 1993. Taylor \& Francis.


\bibitem{neal2003slice}
Neal, Radford M.
\newblock Slice sampling.
\newblock \emph{The annals of statistics}, 31(3):705--767, 2003. Institute of Mathematical Statistics.


\bibitem{rondeau2012frailtypack}
Rondeau, Virginie and Marzroui, Yassin and Gonzalez, Juan R.
\newblock frailtypack: an R package for the analysis of correlated survival data with frailty models using penalized likelihood estimation or parametrical estimation.
\newblock \emph{Journal of statistical Software}, 47:1--28, 2012.


\bibitem{khan2022accelerated}
Khan, Shahedul A and Basharat, Nyla.
\newblock Accelerated failure time models for recurrent event data analysis and joint modeling.
\newblock \emph{Computational Statistics}, 37(4):1569--1597, 2022. Springer.


\bibitem{dearmon2016local}
Dearmon, Jacob and Smith, Tony E.
\newblock Local marginal analysis of spatial data: a Gaussian process regression approach with Bayesian model and kernel averaging.
\newblock \emph{Spatial econometrics: qualitative and limited dependent variables}, 297--342, 2016. Emerald Group Publishing Limited.


\end{thebibliography}
\end{document}